\documentclass{iau}
\usepackage{graphicx}

\title[Plasma Properties and Magnetic Field in a Prominence-like Structure] 
{Estimation of Plasma Properties and Magnetic Field in a Prominence-like Structure as Observed by SDO/AIA}

\author[Dwivedi et al.]   
{B.N. Dwivedi$^1$, A.K. Srivastava$^2$
 \and Anita Mohan$^1$}

\affiliation{$^1$Department of Physics, IIT (BHU), Varanasi, India \\ email: {\tt bnd.app@iitbhu.ac.in} \\[\affilskip]
$^2$Aryabhatta Research Institute of Observational Sciences (ARIES), Nainital, India}

\pubyear{2013}
\volume{xxx}  
\pagerange{}
\jname{IAU Symposium 300 : Nature of prominences and their role in space weather}
\editors{B. Schmieder, J.-M. Melherbe, S. Wu}
\begin{document}

\maketitle

\begin{abstract}
We analyze a prominence-like cool plasma structure as observed by Atmospheric Imaging Assembly (AIA) onboard the Solar Dynamics Observatory (SDO). We perform the Differential Emission Measure (DEM) analysis using various filters of AIA, and also deduce the temperature and density structure in and around the observed flux-tube.  In addition to deducing plasma parameters, we also find an evidence of multiple harmonics of fast magnetoacoustic kink waves in the observed prominence-like magnetic structure. Making use of estimated plasma parameters and observed wave parameters, under the baseline of MHD seismology, we deduce magnetic field in the flux-tube. The wave period ratio P1/P2 = 2.18 is also observed in the flux-tube, which carries the signature of magnetic field divergence where we estimate the tube expansion factor as 1.27. We discuss constraints in the estimation of plasma and magnetic field properties in such a structure in the current observational perspective, which may shed new light on the localized plasma dynamics and heating scenario in the solar atmosphere.
\end{abstract}

\firstsection 
\section{Introduction}

In this brief paper, we report the differential emission measure (DEM) analysis using various AIA filters and deduce  temperature structure of the emitting region using the method of Aschwanden et al. (2013). We also estimate local  magnetic field in the tube from MHD seismology. We have carried out emission measure and temperature distribution analysis as well as tracing and estimating the density along the prominence by extensively using the automated method. SDO/AIA data provide information about the emission measure and estimation of the average density and temperature. We use the full-disk SDO/AIA in all EUV channels around 13:21 UT on 7 March 2011. We calibrate and clean the data using the aia\_prep subroutine of SSWIDL and co-align the AIA images as observed in its various AIA filters using the co-alignment test. We obtain emission measure and temperature maps for six AIA filter full-disk and co-aligned images (304~\AA, 171\AA, 193~\AA, 94~\AA, 335~\AA, and 211~\AA) in the temperature between 0.4 MK and 9.0 MK. The partial FOV analysis in the form of the emission measure and temperature measurements has been performed. The prominence structure is not fully resolved in multi-temperature emissions as it only contributes to the information of 304~\AA~He II emission. However, the cool plasma streaks are evident in the prominence sub-FOV maintained at temperature below 1.0 MK.
\vspace{-0.5cm}
\section{Results}

The main results obtained are briefly described {\bf below}:

[1] The automated emission analyses, using Aschwanden et al. (2013) method, clearly indicate the presence of cool plasma streaks in the prominence structure. The electron density decreases in the highly tangled prominence structure up to its apex which lies at a distance of $\approx$ 85 Mm from the left-footpoint of the flux-tube. It shows some increment towards right foot-point side along the tube. The density near the apex of the prominence is Ne = 1.0$\times$10$^{9}$ cm$^{-3}$. The density near the left foot-point lies in the range of Ne = 3.16$\times$10$^{9}-$1.0$\times$10$^{10}$ cm$^{-3}$.

[2] The temperature estimations with best least chi square show a uniform electron temperature (Te$\sim$5.0$\times$$10^{5}$ K) below 1.0 MK along the prominence-like cool loop system. However, the calculated average temperature is 1.0 MK, because a few temperature points are estimated at $>$ 1.0 MK. The average temperature of about 1.0 MK is two orders of magnitude large than the cool prominence plasma temperatures (below 10$^{4}$ K) and most likely represent a very hot part of the prominence-corona transition region.

[3] Srivastava et al. (2013) observed the multiple harmonics of the fast magnetoacoustic kink waves in this prominence-like cool loop system. 

[4] Srivastava et al. (2013) have found $\approx$ 667 s and $\approx$ 305 s periods respectively near the apex and foot-point of the prominence-like cool loop system. The phase speed of fundamental mode period (within tube length of $\sim$ 170 Mm) lies in the fast regime (510 km s$^{-1}$) of the tube MHD waves, which can either be magnetoacoustic sausage or kink waves. However, under the given morphology and plasma conditions (typical coronal density inside the prominence), sausage modes are unlikely because of longer wavelength cut-off issue. Therefore, the most likely fast tubular mode is the non-linear magneto-acoustic kink waves that can weakly modulate the density near the boundary of the flux-tube (and thus intensity) during the modulation of plasma column depth in the obliquely oriented flux-tube  in accordance with Cooper et al. (2003) theory.

[5] The period ratio shift of the fundamental mode to the first harmonics is $>$ 2.0, which indicates the flux-tube expansion and the magnetic field stratification in the loop (Srivastava et al. 2013). This is dominant over the longitudinal density structuring. For the first time, the expansion factor of the loop was estimated to be 1.27. It was also suggested as the first clues of the shift of fundamental mode period by a factor of 0.85 in the duration of 600 s. This provides an observational clues to further develop the appropriate theory (Morton and Erdélyi, 2009
, Ruderman 2011). 

[6] The average density near the apex of the observed filament is Ne = 1.0 $\times$10$^{9}$ cm$^{-3}$. Therefore, we estimate the magnetic field of the prominence near its apex as 9.0 Gauss based on the theory of MHD seismology and observational signature of the fundamental kink waves (Nakariakov and Ofman, 2001). Therefore, we estimate the density (Ne = 1.0 $\times$10$^{9}$ cm$^{-3}$), temperature ($\sim$ 1.0 MK), and magnetic field (9.0 Gauss) values in the prominence-like flux-tube observed by SDO/AIA.

\vspace{-0.5cm}
\section{Acknowledgements.}
We thank the referee for his/her suggestions. BND’s participation in the IAUS300 is enabled by the financial support from the ESA. We acknowledge SDO/AIA-NASA for data, and E. O’Shea for the randomlet software, and Torrence and Compo for the wavelet software. We also acknowledge the use of the procedures for automated DEM and Te - analyses developed by M.J. Aschwanden. AKS acknowledges the support from DST-RFBR and Indo-Austrian Projects (INT/RFBR/P-117;INT/AUA/BMWF/P-18/2013), and also thanks Shobhna Srivastava for her patient encouragement during research.

\end{document}